\documentclass[twocolumn,showpacs,preprintnumbers,amsmath,amssymb,prc]{revtex4}
\usepackage{graphicx}% Include figure files
\usepackage{dcolumn}% Align table columns on decimal point
\usepackage{bm}% bold math
\begin{document}
\title{Bimodality emerges from transport model calculations of heavy ion
collisions at intermediate energy}
\author{S. Mallik$^1$, S. Das Gupta$^2$ and G. Chaudhuri$^{1}$}

\affiliation{$^1$Theoretical Nuclear Physics Group, Variable Energy Cyclotron Centre, 1/AF Bidhan Nagar, Kolkata 700064, India}
\affiliation{$^2$Physics Department, McGill University, Montr{\'e}al, Canada H3A 2T8}
\begin{abstract}
This work is a continuation of our effort \cite{Mallik1} to examine if signatures of a phase transition can be extracted from transport
model calculations of heavy ion collisions at intermediate energy.  A signature of first order phase transition is the appearance of a
bimodal distribution in $P_m(k)$ in finite systems.  Here $P_m(k)$ is the
probability that the maximum of the multiplicity distribution occurs at mass
number $k$.  Using a well-known model for event generation (BUU plus
fluctuation), we do two cases of central collision: mass 40 on mass 40 and
mass 120 on mass 120.  Bimodality is seen in both the cases.  The results are quite similar to those obtained in statistical model calculations.

An intriguing feature is seen.  We observe that at the energy where bimodality occurs. other phase transition like signatures appear.  There are
breaks in certain first order derivatives.  We then examine if such breaks
appear in standard Botzmann-Uehling-Uhlenbeck (BUU) calculations without
fluctuations.  They do.  The implication is interesting.  If first order phase transition occurs, it may be possible to recognise that from ordinary
BUU calculation.  Probably the reason this was not seen already is because
this aspect was not investigated before.
\end{abstract}
\maketitle
\section{Introduction}
In a recent paper \cite{Mallik1} we used a well-known model of fluctuations \cite{Bauer1} in BUU \cite{Bertsch1} to generate event by event simulation of collisions of fairly large (120 on 120) ions as well as not so large (40 on 40) ions.  The multiplicity distribution of final products of collision showed a remarkable similarity with the results given by equilibrium statistical models where we used a canonical thermodynamic model (CTM) \cite{Das1}. Both canonical \cite{Dasgupta1,Das1} and grand canonical thermodynamic models \cite{Bugaev1} predict first order phase transitions in hot nuclear systems. So the similarity
suggested that probably transport model calculations also will give more direct evidence of first order phase transition.  This work is aimed at exploring this further.  It is not so obvious how to go about doing this. In canonical and grand canonical models there are two parameters, temperature $T$ (which is the basic parameter) and average energy $E$. The behaviour of $E$ against $T$ can indicate the order of phase transition. Usually two parameters are needed but in transport model calculations that we do here there is only one parameter, the beam energy $E$.  Defining a temperature is quite difficult.  Formulae like $\frac{E*}{A}=\frac{3T}{2}$ are obviously inappropriate.  One might try $T=(\frac{\partial S}{\partial E})_V$ but that requires obtaining the entropy of
an interacting system and an accurate evaluation would be very hard.

We recall that as early as in 1998, compiling existing knowledge from experimental data and comparing these with lattice gas model predictions, it was concluded that in intermediate energy heavy ion collisions one passes through a first order phase transition \cite{Pan1}.  This was subsequently investigated by many authors with different approaches.  One approach uses the idea of ``bimodality''  A very useful exposition of this can be found in \cite{Gulminelli1} which also has a list of other references using ``bimodality'' approach.  The size of the largest cluster $k$ is considered to be an order parameter.
Phase transitions occur in very large systems  but practical calculations (and experiments with heavy ions) need to be done with  finite systems. Gulminelli and Chomaz point out that we should expect for $P_m(k)$ (probability that the biggest cluster has mass $k$) a double humped distribution (hence the name bimodality) if the phase transition is first order.  The authors establish this with a lattice gas model.  For relevant study of this in Ising model, see \cite{Binder1}  Bimodality also emerges in CTM which has a first order phase transition.  This was studied in \cite{Gargi1} and we will have occasion to return to bimodality in CTM later. The objective of this work is to investigate if bimodality emerges from a transport model calculation.  Using what is labelled as QMD (Quantum Molecular Dynamics), Lefevre and Aichelin have used ideas from bimodality to show that in some non-central collisions \cite{Pichon1} there is evidence of first order phase transition \cite{Fevre1}.  In the calculation the full distribution $P_m(k)$ was not displayed.  The
complete curves $P_m(k)$ as a function of beam energy is quite interesting and we present them here. In contrast with the QMD work we use central collisions. As we are interested in phase transition under the influence of nuclear force, Coulomb effects will be switched off. The use of central collisions to display bimodality has been questioned before.  Also the transport model we use is quite different from QMD.
\section{The Model}
The calculations done here follow those of \cite{Mallik1} except for small
but important details which will be fully presented.  For completeness
we outline the model.  More details are given in \cite{Mallik1}.
The original model was developed in \cite{Bauer1} where the formal structure
was discussed and an application was presented.
Initially each nucleon in the target and the projectile is given a
semi-classical phase
space density.  For each nucleon this phase space density is represented by
$\tilde{N}$ test particles where each test particle is generated by Monte-Carlo
and  has a position $\vec r$ and a momentum $\vec p$.  Initially the two
nuclei are apart with an impact parameter $b$ (in this work we only consider
central collision $b$=0) and the projectile starts with a beam velocity
towards the target.  As they propagate in time, the test particles will
move in a mean field and suffer hard scattering.  As $\tilde{N}$ test particles
will represent a nucleon
the collision cross-section between test particles is reduced to
$\sigma_{nn}/\tilde{N}$ where $\sigma_{nn}$ is nucleon-nucleon cross-section.
In \cite{Bauer1}, to simulate an event,
the cross-section is further reduced by a factor $\tilde{N}$ but if a collision
happens not only the the two test particles go from $\vec {p_1}$ to
$\vec {p_1}+\Delta \vec {p}$ and from $\vec {p_2}$ to
$\vec{p_2}-\Delta \vec{p}$ but $\tilde{N}-1$ test particles contiguous to
test particle 1 undergo momentum change $\Delta \vec {p}$ and $\tilde{N}-1$
test particles contiguous to test particle 2 undergo momentum change
$-\Delta \vec{p}$.  This is followed in time till the collisions are over
and we have one event.  To simulate another event we start with initial
positions of the ions and generate by Monte-carlo fresh sets of test particles.
Many events are needed before any comparison with experiments can be made.

The calculation for each event is quite large as collisions between
$(A_p+A_t)\tilde{N}$ test particles need to be checked.  Here $A_p$ is
the number of nucleons in the projectile and $A_t$ is the number of nucleons
in the target and $\tilde{N}$ is rather large (usually about 100).
It was shown in
\cite{Mallik1} that the problem can be reduced, for each event, to checking
collisions between just $(A_p+A_t)$ test particles.  This feature makes it
possible for us to do large systems.
We refer to section II of \cite{Mallik1} for details.  No compromise to
theory or numerical accuracy is introduced.
\begin{figure}[h]
\begin{center}
\includegraphics[width=5cm,keepaspectratio=true]{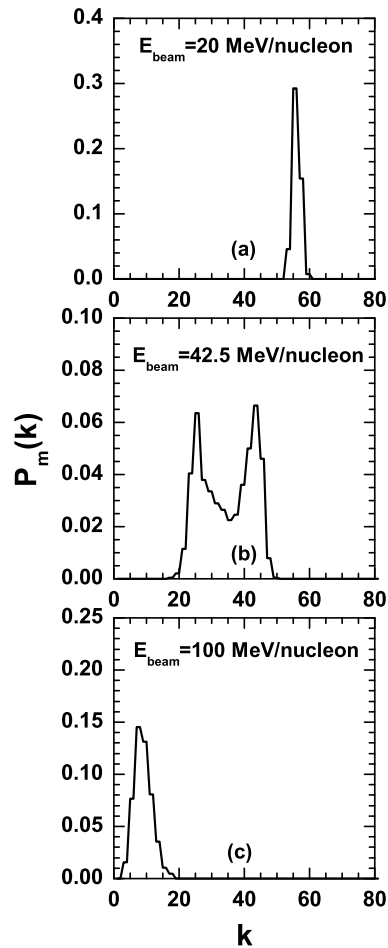}
\caption{ Largest cluster probability distribution for $A_p=40$ on $A_t=40$ reaction at beam energies (a)20 MeV/nucleon, (b)42.5 MeV/nucleon and (c)100 MeV/nucleon.  The average value of 2 mass units are shown. At each energy 1000 events are chosen. The results shown in this figure are calculated at
$t$=300 $fm/c$}
\end{center}
\label{Bimodality_smaller_system}
\end{figure}

\section{Some details of the simulation}
For completeness, we provide some details of the calculation that will
be needed to explain our cluster recognition algorithm. Collisions
are treated as in \cite{Bertsch1}
For Vlasov
propagation we use the lattice Hamiltonian method \cite{Lenk1} which
accurately conserves energy and momentum. The mean field is also taken from \cite{Lenk1}. The configuration space is divided
into cubic lattices.  The lattice points are $l$ fm apart.  Thus the
configuration space is discretized into boxes of size $l^3$ $fm^3$.  Density
at the lattice point $\vec{r_{\alpha}}$ is given by
\begin{equation}
\rho_L(\vec{r_{\alpha}})=\sum_iS(\vec{r_{\alpha}}-\vec{r_i})
\end{equation}
\begin{figure}[t]
\begin{center}
\includegraphics[width=5cm,keepaspectratio=true]{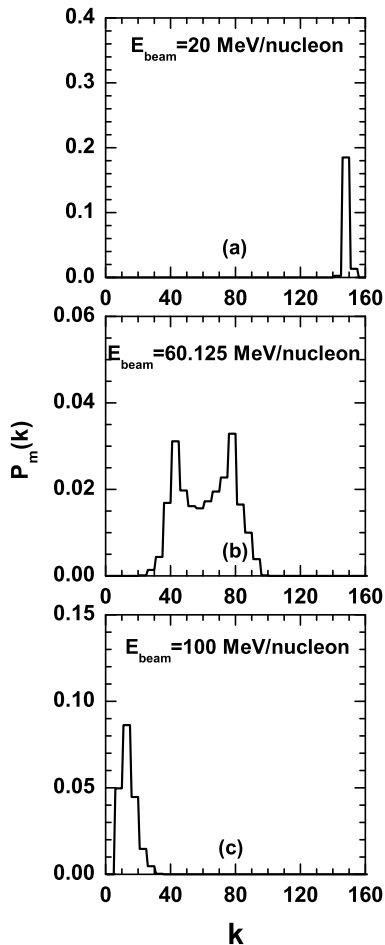}
\caption{ Largest cluster probability distribution for $A_p=120$ on $A_t=120$ reaction at beam energies (a)20 MeV/nucleon, (b)60.125 MeV/nucleon and (c)100 MeV/nucleon.  The average value of 5 mass units are shown. At each energy 1000 events are chosen. The results shown here are calculated at $t$=600 $fm/c$.}
\end{center}
\label{Bimodality_larger_system}
\end{figure}

\begin{figure}[h]
\begin{center}
\includegraphics[width=\columnwidth,keepaspectratio=true]{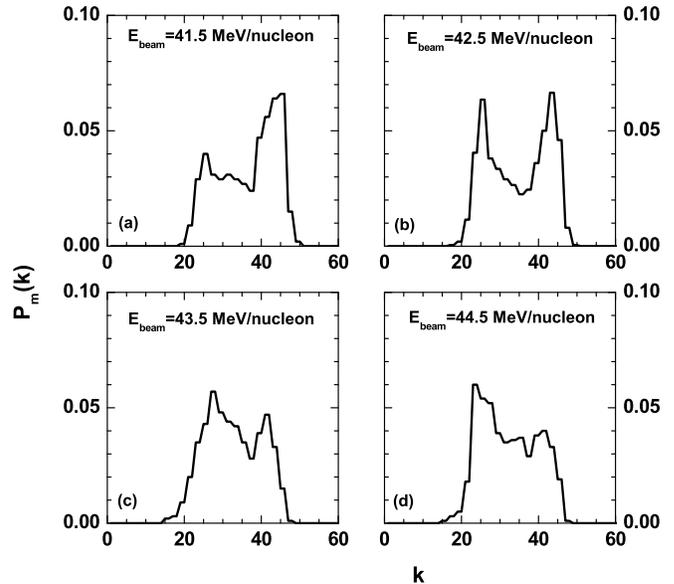}
\caption{ Largest cluster probability distribution for $A_p=40$ on $A_t=40$ reaction at beam energies (a)41.5 MeV/nucleon, (b)42.5 MeV/nucleon, (c)43.5 MeV/nucleon and (d)44.5 MeV/nucleon.  The average value of 2 mass units are shown. At each energy 1000 events are chosen. The results shown in this figure are calculated at $t$=300 $fm/c$}
\end{center}
\label{Bimodality_near_transition}
\end{figure}
Here the sum over $i$ goes over all the test particles and the form factor is
\begin{equation}
S(\vec {r})=\frac{1}{\tilde{N}(nl)^6}g(x)g(y)g(z)
\end{equation}
where
\begin{equation}
g(q)=(nl-|q|)\Theta(nl-|q|)
\end{equation}
In this work we have used $n=1$ and $l=1$ $fm$.  Because of this choice, at a given time, if two test particles are more than 2 $fm$ apart,
they can not affect each other's motion directly.  This prompts us to prescribe the following algorithm.  Two test particles are part of
the same cluster if the distance between them is less than or equal to 2 $fm$.  Two clusters are distinct if none of the test particles
of cluster 1 is within a distance of 2 $fm$ from any of the test particles of cluster 2.  With this prescription, the number of clusters and
their sizes will change as a function of time at early times.  Because of momenta that test particles carry, two test particles which
were less than 2 $fm$ apart (or more than 2 $fm$ apart) may not remain so at a later time.  The physical picture we depend upon is that
two heavy ions collide, clusters are formed which begin to move away from one another.  If this is true then at large times, the momentum $\vec{p_i}$
and position $\vec{r_i}$ in each individual cluster are strongly correlated and transfer of test particles  between different clusters will disappear.
One can test this by plotting the multiplicity distribution as a function of time.  We find that for 40 on 40 near constancy is observed
around 300 $fm/c$ and for 120 on 120 (because this is a much larger system) around 600 $fm/c$.  From the multiplicity distributions of
1000 events, we construct $P_m(k)$, the probability that the largest cluster in an event has $k$ nucleons.  Examples are shown in Figs. 1 and 2.

Our algorithm for enumerating cluster numbers and their sizes has some similarities and also some differences with the method used in \cite{Suneel}
in QMD.

\begin{figure}[h]
\begin{center}
\includegraphics[width=6cm,keepaspectratio=true]{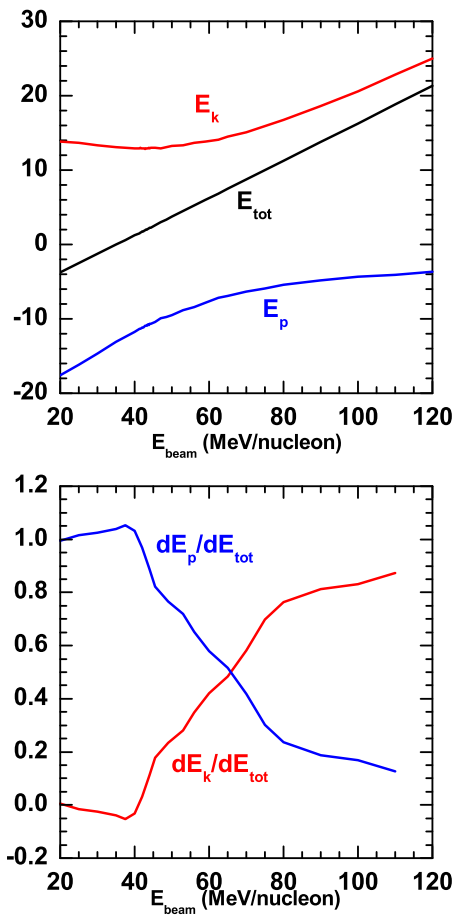}
\caption{ (Color online) Upper panel: Dependence of kinetic energy per nucleon (red), potential  energy per nucleon (blue) and total  energy per nucleon (black) for $A_p=40$ on $A_t=40$ reaction on the projectile beam energy per nucleon.\\
Lower panel: Dependence of first order derivatives of kinetic energy and potential energy with respect to total energy on total enrgy per nucleon for $A_p=40$ on $A_t=40$ reaction.}
\end{center}
\label{Energy_smaller_system}
\end{figure}
\section{Results}
In order to study bimodality from our event generation model (BUU plus fluctuation) we simulate central collisions of mass $40$ on mass $40$ and mass $120$ on mass $120$ at different projectile beam energies. For $40$ on $40$ reaction the largest cluster probability distribution is plotted in Fig. 1 for $E_{beam}=20$, $42.5$ and $100$ MeV/nucleon. At each energy $1000$ events are taken and for each event calculation is done up to $t=300$ fm/c. The results shown are averages for graphs of 2 consecutive mass number at $t$=300 fm/c. At projectile beam energy ($E_{beam}$)=20 MeV/nucleon, the $P_m(k)$ is peaked at around mass 60 which represents liquid phase where as at $E_{beam}$=100 MeV/nucleon, the probability distribution peaks at very low mass i.e. it suggests the system is in the gas phase. In between these two extremes, at $E_{beam}$=42.5 MeV/nucleon the largest cluster probability distribution shows the bimodal behaviour where the height of the two peaks are almost same.
Fig. 2 shows similar features for for much heavier system :120 on 120.
Here we take the results at 600 $fm/c$. Several points are worth mentioning.  Whether in the case of 40 on 40 or 120 on 120 bimodality occurs in a very narrow range of energy.  For 40 on 40 we demonstrate that in Fig.3.  Thus to locate bimodality in experiments beam energy variation has to be done
in small energy steps.  The narrow width of energy over which bimodality appears is common in CTM also.\\
A phase transition like behaviour emerges more directly from our calculations.  This is quite revealing.  For 40 on 40 (and 120 on 120) we do our
calculation as a function of beam energy.  For example for 40 on 40 we did our calculation from beam energy 20 MeV/nucleon to
100 MeV/nucleon.  For each beam energy 1000 events were generated.  From these events we compute the average total energy $E_{tot}$, the average kinetic energy
$E_k$ and the average potential energy $E_p$ per particle.  let us plot the total energy $E_{tot}$ in the cm.  This will of course increase in value as $E_{beam}$ (MeV/nucleon) increases.  This energy $E_{tot}$ is the sum of kinetic energy $E_k$ and potential energy $E_p$.  The origin of $E_k$ is more complicated.  It arises from Fermi motion of the test particles and also the cm kinetic energy of each cluster.  The quantity $E_p$ is more straightforward.  It arises from the potential energy of the clusters.  An insight is obtained by examining the derivative $dE_p/dE_{tot}$.  A sudden change in the derivative $dE_p/dE_{tot}$ occurs at the point where bimodality is observed.  This type break in the first derivative is typical of first order phase transitions.  We might consider this break to be an additional signature of a first order phase transition.\\
Since here we plotted values for the average of many events, it is natural to ask: could it not be seen in standard BUU which does give average values.  This is not obviously so as the average might depend also on the details of fluctuations that were used in our event generation model.  However straightforward BUU as has been used before \cite{Bertsch1} does produce similar result (Fig.6).  Thus the possibility exists that one might get signature for first order phse transition from
BUU itself.

\begin{figure}[h]
\begin{center}
\includegraphics[width=6cm,keepaspectratio=true]{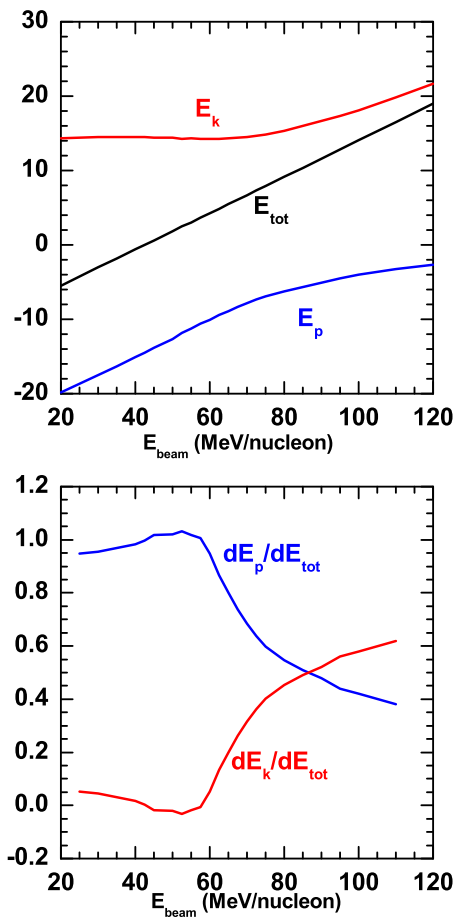}
\caption{ (Color online) Same as Fig. 3 but here the nuclear reaction is $A_p=120$ on $A_t=120$ instead of $40$ on $40$.}
\end{center}
\label{Energy_larger_system}
\end{figure}
\section{Discussion}
We have done central collisions of 40 on 40 and 120 on 120 to test appearance of bimodality which is considered to be a signature of first order phase transition in finite systems.  Bimodality was observed.  Since calculations were done with fixed beam energy one might be tempted to call it a microcanonical calculation.  But even in central collisions  at least two different reaction mechanisms operate.  One is
collision between peripheral parts.  Here some nucleons may simply pass by or at most make one collision.  We would include pre-equilibrium emission in this category.  The number of nucleons in pre-equilibrium emission and the energy they carry off will vary from event to event.  Thus the number of nucleons which suffer multiple collisions and the energy that is available for such multiple collisions event will vary.  Presumably such multiple collision events can show signatures statistical equilibrium, phase transitions etc but in experiments and in transport model calculations such as this one all different reaction mechanisms will play a role.  Nonetheless, this calculation shows that with just nuclear forces first order phase transition is possible in intermediate energy heavy ion collisions.\\
\begin{figure}[t]
\begin{center}
\includegraphics[width=6cm,keepaspectratio=true]{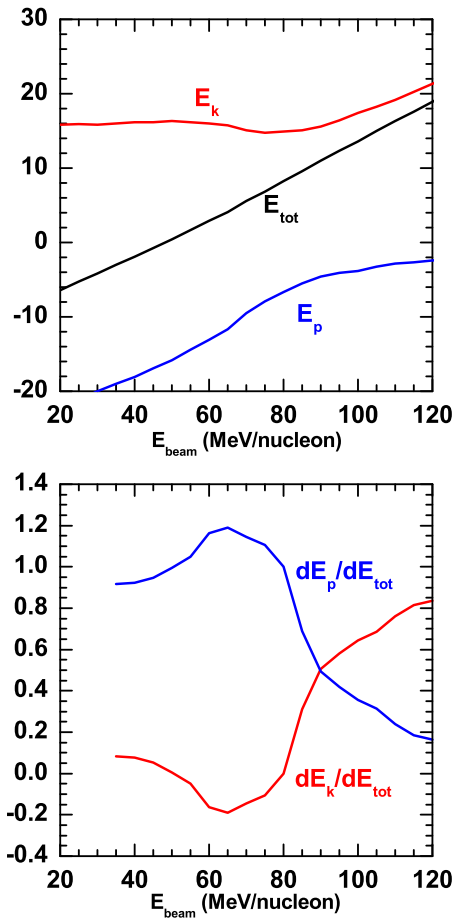}
\caption{ (Color online) Same as Fig. 5 but here the calculation is done by standard BUU method (i.e. without fluctuation).}
\end{center}
\label{Energy_larger_system_normal_BUU}
\end{figure}
The closeness of CTM results and transport model results might be exploited to estimate the freeze-out density in statistical models. In CTM a freeze-out density is assumed but there is no such parameter in transport model. In CTM the temperature at which bimodality appears depends on the assumed freeze-out density. There will be a freeze-out density at which CTM gives the same bimodality temperature as the transport model. This could be an estimate for freeze-out density. Detailed calculations have not been carried out.

\end{document}